\documentclass[oneside,11pt]{article}

\usepackage[utf8]{inputenc}
\usepackage[T1]{fontenc}

\usepackage{amsthm}
\usepackage{amsmath}
\usepackage{amsfonts}
\usepackage{amssymb}

\usepackage{graphicx} 
\usepackage{float}
\usepackage{booktabs}
\usepackage{multirow}
\usepackage{rotating}
\usepackage{array}
\usepackage{xcolor}

\usepackage{nowidow}

\usepackage{breakcites}

\usepackage{enumitem}

\usepackage[top=1.5cm,bottom=2cm,right=2.5cm,left=2.5cm]{geometry}
\usepackage{titling}

\usepackage{natbib}

\usepackage{ifplatform}

\usepackage{textcomp}

\ifwindows
\fi

\ifwindows
\usepackage{bbm}
\else
\iflinux
\usepackage{bbm}
\else
\usepackage{bbm}
\fi
\fi

\newcommand{\lp}{\left(}
\newcommand{\rp}{\right)}
\newcommand{\lc}{\left[}
\newcommand{\rc}{\right]}

\newcommand{\R}{\mathbb{R}}

\newcommand{\bx}{\mathbf{x}}
\newcommand{\be}{\mathbf{e}}

\newcommand{\by}{\mathbf{y}}
\newcommand{\bX}{\mathbf{X}}
\newcommand{\bY}{\mathbf{Y}}

\newcommand{\bz}{\mathbf{z}}

\newcommand{\zero}{\mathbf{0}}

\newcommand{\bbe}{\boldsymbol\beta}

\newcommand{\bthe}{\boldsymbol\theta}
\newcommand{\btheta}{\boldsymbol\theta}

\newcommand{\X}{\boldsymbol{\mathcal{X}}_{\bx,p}}
\newcommand{\W}{\boldsymbol{\mathcal{W}}_{\bx}}

\newcommand{\bB}{\mathbf{B}}

\newcommand{\bI}{\mathbf{I}}

\newcommand{\lrp}[1]{\left(#1\right)}

\newcommand{\lrb}[1]{\left\{#1\right\}}

\newcommand{\Esp}[1]{\mathbb{E}\lc #1\rc}
\newcommand{\V}[1]{\mathbb{V}\mathrm{ar}\lc #1\rc}

\newcommand{\diag}[1]{\mathrm{diag}\lp #1\rp}

\newcommand{\norm}[1]{\left|\left| #1\right|\right|}

\newcommand{\Om}[1]{\Omega_{#1}}
\newcommand{\om}[1]{\omega_{#1}}
\newcommand{\Iq}[2]{\int_{\Omega_{q}} #1\,\omega_{q}(d #2)}
\newcommand{\Iqr}[3]{\int_{\Omega_{q}\times\R} #1\,d #3\,\omega_{q}(d #2)}

\newcommand{\Ir}[2]{\int_{\R} #1\,d #2}
\newcommand{\Iqq}[3]{\int_{\Omega_{q_1}\times\Omega_{q_2}} #1\,\omega_{q_2}(d #3)\,\omega_{q_1}(d #2)}

\DeclareFontFamily{OT1}{pzc}{}
\DeclareFontShape{OT1}{pzc}{m}{it}{<-> s * [1.10] pzcmi7t}{}
\DeclareMathAlphabet{\mathpzc}{OT1}{pzc}{m}{it}

\usepackage[pdftex,final,bookmarksnumbered,bookmarksopen=false,breaklinks,colorlinks]{hyperref}
\hypersetup{
	final,
	bookmarksnumbered=true,
	bookmarksopen=true,
	bookmarksopenlevel=0,
	unicode=false,          
	pdftoolbar=true,        
	pdfmenubar=true,        
	pdffitwindow=false,     
	pdftitle={Smoothing-based tests with directional random variables},    
	pdfdisplaydoctitle=true, 
	pdftoolbar=true,
	pdfmenubar=true,
	pdflang={English},
	pdfauthor={Eduardo Garcia-Portugues, Rosa M. Crujeiras, Wenceslao Gonzalez-Manteiga},     
	pdfsubject={arXiv paper},   
	pdfcreator={Eduardo Garcia-Portugues},   
	pdfproducer={Eduardo Garcia-Portugues}, 
	pdfkeywords={Directional data}{Goodness-of-fit}{Kernel smoothing}{Tests}, 
	pdfnewwindow=true,      
	breaklinks=true,
	hidelinks,       
	linkcolor=black,          
	citecolor=black,        
	filecolor=magenta,      
	urlcolor=cyan           
}

\allowdisplaybreaks

\begin{document}

\title{Smoothing-based tests with directional random variables}
\setlength{\droptitle}{-1cm}
\predate{}%
\postdate{}%

\date{}

\author{Eduardo Garc\'ia-Portugu\'es$^{1,2,4}$, Rosa M. Crujeiras$^3$, and Wenceslao Gonz\'alez-Manteiga$^3$}

\footnotetext[1]{
	Department of Statistics, Carlos III University of Madrid (Spain).}
\footnotetext[2]{
	UC3M-BS Institute of Financial Big Data, Carlos III University of Madrid (Spain).}
\footnotetext[3]{
	Department of Statistics, Mathematical Analysis and Optimization, University of Santiago de Compostela (Spain).}
\footnotetext[4]{Corresponding author. e-mail: \href{mailto:edgarcia@est-econ.uc3m.es}{edgarcia@est-econ.uc3m.es}.}

\maketitle


\begin{abstract}
	Testing procedures for assessing specific parametric model forms, or for checking the plausibility of simplifying assumptions, play a central role in the mathematical treatment of the uncertain. No \emph{certain} answers are obtained by testing methods, but at least the \emph{uncertainty} of these answers is properly quantified. This is the case for tests designed on the two most general data generating mechanisms in practice: distribution/density and regression models. Testing proposals are usually formulated on the Euclidean space, but important challenges arise in non-Euclidean settings, such as when directional variables (\textit{i.e.}, random vectors on the hypersphere) are involved. This work reviews some of the smoothing-based testing procedures for density and regression models that comprise directional variables. The asymptotic distributions of the revised proposals are presented, jointly with some numerical illustrations justifying the need of employing resampling mechanisms for effective test calibration.
\end{abstract}
\begin{flushleft}
\end{flushleft}

\section{On goodness-of-fit tests and smoothing}
\label{gpcgm:sec:intro}

In the early years of the 20th century, K. Pearson and colleagues initiate the development of testing methods for assessing the goodness-of-fit of a certain parametric model. \cite{Pearson1900} presents his celebrated $\chi^2$ test as a criterion to check if a given \emph{system of deviations} from a theoretical distribution could be supposed to come from random sampling, but it is not until a couple of years later when \cite{Elderton1902} coined the term goodness-of-fit \emph{of theory to observation}. Also at the beginning of last century, \cite{Pearson1916} introduce the first ideas for goodness-of-fit tests in regression models. With no theoretical support from probability theory (which was developed almost at the same time, and therefore, its impact on statistics was noticed some years later), these works set the basis for the construction of testing procedures with the aim of assessing a certain parametric null hypothesis for density/distribution (see \cite{Bickel1973} and \cite{Durbin1973}, as two influential papers) and regression models (see \cite{Gonzalez-Manteiga2013} for a complete review on goodness-of-fit tests in this setting). \\

This work focus on a certain class of tests that makes use of nonparametric (smooth) estimators of the target function, that is, the density or the regression functions. First, consider the problem of testing a certain parametric density model
\begin{align}
H_0:\;f\in\mathcal F_{\Theta}\quad\mbox{vs.}\quad H_1:\;f\notin\mathcal F_{\Theta},
\label{gpcgm:eq:testdensity}
\end{align}
with $\mathcal{F}_{\Theta}=\{f_\theta:\theta\in\Theta\}$ a parametric density family. From a smoothing-based perspective, a pilot estimator $\hat f$ constructed from $X_1,\ldots,X_n$, a sample from the random variable (rv) $X$, will be confronted with a parametric estimator by the use of a certain discrepancy measure. \cite{Bickel1973} consider the classical Kernel Density Estimator (KDE) $\hat f_g(x)=\frac{1}{ng}\sum_{i=1}^nK\big(\frac{x-X_i}{g}\big)$, with kernel $K$ and bandwidth $g$, to be compared with a parametric estimator $f_{\hat\btheta}$ under the null through an $L^2$-distance. In general, test statistics for \eqref{gpcgm:eq:testdensity} can be built as $T_n=d(\hat f,f_{\hat\btheta})$, being $d$ a discrepancy measure between both estimators. \\

The ideas of goodness-of-fit tests for density curves have been naturally extended in the nineties of the last century to regression models. Consider, as a reference, a regression model $Y=m(X)+\varepsilon$, where the goal is to test
\begin{align}
H_0:\; m\in \mathcal{M}_\Theta\quad\text{vs.}\quad H_1:\; m\notin \mathcal{M}_\Theta
\label{gpcgm:eq:testregression}
\end{align}
in an omnibus way from a sample $\lrb{\lrp{X_i,Y_i}}_{i=1}^n$ of $(X,Y)$. Here $m(x)=\Esp{Y\vert X=x}$ is the regression function of $Y$ over $X$, and $\varepsilon$ is a random error such that $\Esp{\varepsilon\vert X}=0$. A pilot estimator $\hat m(x)=\sum_{i=1}^n W_{n,i}(x)Y_i$ can be constructed using nonparametric weights, such as the Nadaraya--Watson weights given by $W_{n,i}(x)=K\big(\frac{x-X_i}{g}\big)\big/\sum_{j=1}^n K\big(\frac{x-X_j}{g}\big)$. Other possible weights, such as the ones from local linear estimation, $k$-nearest neighbours, or splines, can be also considered. Using these kind of pilot estimators, tests statistics can be built (similarly to the density case) as $T_{n}=d\lrp{\hat m,m_{\hat\btheta}}$. In the presence of directional random variables, and considering the previous smoothing ideas, similar tests can be developed.

\section{Goodness-of-fit tests with directional data}
\label{gpcgm:sec:gofdir}

The statistical analysis of \textit{directional data}, this is, elements in the $q$-sphere $\Omega_{q}=\{\bx\in\R^{q+1}:\bx'\bx=1\}$, is notably different from the analysis of \textit{linear} (Euclidean) data. In particular, no canonical ordering exists in $\Omega_{q}$, which makes rank-based inference ill-defined. We refer to the book of \cite{Mardia2000} for a comprehensive treatment of statistical inference with directional data, and for a collection of applications. Some smooth estimators for density and regression in this context are briefly revised below. These estimators are used as pilots for the testing proposals introduced in the subsequent sections.

\subsection{Smooth estimation of density and regression}

Let $\bX_1,\ldots,\bX_n$ denote a sample from the directional rv $\bX$ with density $f$. \cite{Hall1987} and \cite{Bai1988}\footnote{\cite{Hall1987}'s (1.3) is equivalent to \cite{Bai1988}'s (1.3), but the latter employs a notation with a more direct connection with the usual KDE.} introduce a KDE for directional data, which is defined as follows:
\begin{align}
\hat f_{h}(\bx)=\frac{1}{n}\sum_{i=1}^n L_h(\bx,\bX_i),\quad L_h(\bx,\bX_i)=\frac{c_{h,q}(L)}{n}L\lrp{\frac{1-\bx'\bX_i}{h^2}},\label{gpcgm:eq:kde}
\end{align}
with $L:\R_0^+\rightarrow\R_0^+$ being the kernel, $h>0$ the bandwidth parameter, and 
\begin{align*}
c_{h,q}(L)^{-1}=\lambda_{h,q}(L) h^{q},\quad \lambda_{h,q}(L)=\om{q-1}\int_0^{2h^{-2}} L(r) r^{\frac{q}{2}-1}(2-rh^2)^{\frac{q}{2}-1}\,dr,
\end{align*}
with $\lim_{h\to0}\lambda_{h,q}(L)=\lambda_q(L)=2^{\frac{q}{2}-1}\om{q-1}\allowbreak\int_0^{\infty} L(r) r^{\frac{q}{2}-1}\,dr$. $\om{q}$ denotes both the area of $\Om{q}$, $\om{q}=2\pi^\frac{q+1}{2}/\Gamma\big(\frac{q+1}{2}\big)$, and the Lebesgue measure in $\Om{q}$. For the consistency of \eqref{gpcgm:eq:kde}, it is required that $h=h_n\to0$ when $n\to\infty$ at a rate slower than $nh^q\to\infty$. \\

A directional rv usually appears related to another linear or directional rv, being cylindrical and toroidal data the most common situations in practice. In these scenarios, the modelling approach can be focused on the estimation of the joint density or the regression function. From the first perspective, in order to estimate the density of a directional-linear rv $(\bX,Y)$ in $\Om{q}\times\R$, \cite{Garcia-Portugues:dirlin} propose a KDE adapted to this setting:
\begin{align}	
	\hat f_{h,g}(\bx,y)&=\frac{1}{n}\sum_{i=1}^nLK_{h,g}\lrp{(\bx,y), (\bX_i,Y_i)},\label{gpcgm:eq:kdedl}
\end{align}
where $LK_{h,g}\lrp{(\bx,y), (\bX_i,Y_i)}=L_{h}\lrp{\bx,\bX_i}\times\frac{1}{g}K\lrp{\frac{y-Y_i}{g}}$ is a directional-linear product kernel, and $h,g$ are two bandwidth sequences such that, for the consistency of \eqref{gpcgm:eq:kdedl}, $h,g\to0$ and $nh^qg\to\infty$. \\

In a toroidal scenario, a directional--directional KDE for the density of a rv $(\bX_1,\bX_2)$ in $\Om{q_1}\times\Om{q_2}$ can be derived adapting \eqref{gpcgm:eq:kdedl}:
\begin{align}
\hat f_{h_1,h_2}(\bx_1,\bx_2)=\frac{1}{n}\sum_{i=1}^nLL_{h_1,h_2}\lrp{(\bx_1,\bx_2), (\bX_{1i},\bX_{2i})},\label{gpcgm:eq:kdedd}
\end{align}
with $LL_{h_1,h_2}\lrp{(\bx_1,\bx_2), (\bX_{1i},\bX_{2i})}=L_{h_1}\lrp{\bx_1,\bX_{1i}}\times L_{h_1}\lrp{\bx_2,\bX_{2i}}$, with $h_1,h_2\to0$ and $nh_1^{q_1}h_2^{q_2}\to\infty$ required for consistency. \\

Considering now a regression setting with scalar response and directional covariate, let $\{(\bX_i,Y_i)\}_{i=1}^n$ be a sample from the regression model $Y=m(\bX)+\varepsilon$, where $m(\bx)=\Esp{Y\vert \bX=\bx}:\Om{q}\rightarrow\R$ is the regression function of $Y$ over $\bX$, and $\varepsilon$ is a random error such that $\Esp{\varepsilon\vert \bX}=0$. A nonparametric estimator for $m$, following the \emph{local linear} ideas (see \cite{Fan1996}), can be constructed as follows. Consider a Taylor expansion in a vicinity of $\bX_i$:
\begin{align}
m(\bX_i)\approx m(\bx)+\boldsymbol\nabla m(\bx)'(\bI_{q+1}-\bx\bx')(\bX_i-\bx)=\beta_0+\bbe_1' \bB_{\bx}'(\bX_i-\bx),\label{gpcgm:eq:expans}
\end{align}
where $\bB_{\bx}'\bB_{\bx}=\bI_q$, $\bB_{\bx}\bB_{\bx}'=\bI_{q+1}-\bx\bx'$, and $\bI_q$ is the identity matrix of dimension $q$. From the extension of $m$ to $\bx\in\R^{q+1}\backslash\{\zero\}$ by $m(\bx/\norm{\bx})$, since $\boldsymbol\nabla m(\bx)'\bx=0$, the central expression in \eqref{gpcgm:eq:expans} follows. This motivates the weighted least squares problem
\begin{align}
\min_{(\beta_0,\bbe_1)\in\R^{q+1}}\sum_{i=1}^n \Big(Y_i-\beta_0-\delta_{p,1}\bbe_1'\bB_{\bx}'(\bX_{i}-\bx)\Big)^2 L_h(\bx,\bX_i),\label{gpcgm:eq:wlsp}
\end{align}
where $\delta_{r,s}$ is Kronecker delta, used to control both the local constant ($p=0$) and local linear ($p=1$) fits. The estimate $\hat{\beta_0}$ solving \eqref{gpcgm:eq:wlsp} provides a local linear estimator for $m$:
\begin{align}
\hat m_{h,p}(\bx)=\sum_{i=1}^n W_{n,i}^p(\bx)Y_i,\quad W_{n,i}^p(\bx)=\be_1'\big(\X'\W\X\big)^{-1}\X'\W\be_i,\label{gpcgm:eq:mhp}
\end{align}
where $\bY=(Y_1\,\ldots,Y_n)'$, $\W=\diag{L_h(\bx,\bX_1),\ldots,L_h(\bx,\bX_n)}$, $\be_i$ is the $i$-th unit canonical vector, and $\boldsymbol{\mathcal{X}}_{\bx,1}$ is the $n\times (q+1)$ matrix with the $i$-th row given by $(1,(\bX_i-\bx)'\bB_{\bx})$ (if $p=0$, $\boldsymbol{\mathcal{X}}_{\bx,0}=(1,\ldots,1)'$). For the consistency of \eqref{gpcgm:eq:mhp}, $h\to0$ and $nh^q\to\infty$ are required.

\subsection{Density-based tests}

Testing \eqref{gpcgm:eq:testdensity} allows to check whether there are significant evidences against assuming the density has a given parametric nature, $f_{\btheta_0}$, with parameter $\btheta_0$ either specified (simple hypothesis) or unspecified (composite hypothesis). In the spirit of \cite{Fan1994}'s test, \cite{Boente2013} propose the next test statistic for addressing \eqref{gpcgm:eq:testdensity}:
\[
T_{n,1}=\Iq{\lrp{\hat f_h(\bx)-L_hf_{\hat{\btheta}}(\bx)}^2}{\bx},
\]
where  $L_hf_{\btheta_0}(\bx)=\Iq{L_h(\bx,\by)\allowbreak f_{\btheta_0}(\by)}{\by}=\mathbb{E}_{f_{\btheta_0}}\big[\hat f_h(\bx)\big]$ is the expectation of \eqref{gpcgm:eq:kde} under $f_{\btheta_0}$. This term is included in order to match the asymptotic biases of the nonparametric and parametric estimators. \\

The asymptotic distribution of $T_{n,1}$ is settled on \cite{Zhao2001}'s central limit theorem for the integrated squared error of \eqref{gpcgm:eq:kde}, $I_n=\Iq{(\hat f(\bx)-f(\bx))^2}{\bx}$. The result is given under three different rates for $h\to0$. The relevant one for $T_{n,1}$ is $nh^{q+4}\to0$, when the integrated variance dominates the integrated bias (not dominant under $H_0$), and is given next:
\[
nh^\frac{q}{2}(I_n-\mathbb{E}[I_n])\stackrel{d}{\longrightarrow}\mathcal{N}\lrp{0,2\nu_d^2R(f)},
\]
with $R(f)=\Iq{f(\bx)^2}{\bx}$ (the functional $R(\cdot)$ denotes the integration of the squared argument on its domain of definition) and
\begin{align*}
\nu_d^2=&\,\gamma_q \lambda_q(L)^{-4} \int_0^{\infty} r^{\frac{q}{2}-1}\lrb{\int_0^{\infty} \rho^{\frac{q}{2}-1} L(\rho) \varphi_q(r,\rho) \,d\rho}^2\,dr,\\
\varphi_q(r,\rho)=&\,\left\{ 
\begin{array}{ll}
L\big(r+\rho-2(r\rho)^{\frac{1}{2}}\big)+L\big(r+\rho+2(r\rho)^{\frac{1}{2}}\big), & q=1,\\
\int_{-1}^1 \left(1-\theta^2 \right)^{\frac{q-3}{2}} L\big(r+\rho-2\theta(r\rho)^{\frac{1}{2}}\big)\,d\theta, & q\geq2,\\
\end{array}
\right.\\
\gamma_q=&\,\left\{ 
\begin{array}{ll}
2^{-\frac{1}{2}}, & q=1,\\
\om{q-1}\om{q-2}^2 2^{\frac{3q}{2}-3}, & q\geq2.\\
\end{array}
\right.
\end{align*}
Under certain regularity conditions on $f_{\btheta_0}$ and $L$ (A1--A3 in \cite{Boente2013}), if $\hat\btheta-\btheta_0=\mathcal{O}_{\mathbb{P}}\big(n^{-\frac{1}{2}}\big)$ under $H_0$, then
\[
nh^\frac{q}{2}\lrp{T_{n,1}-\frac{\lambda_q(L^2)\lambda_q(L)^{-2}}{nh^q}}\stackrel{d}{\longrightarrow} \mathcal{N}\lrp{0,2\nu_d^2R(f_{\btheta_0})}.
\]
Hence, asymptotically, the test rejects $H_0$ at level $\alpha$ whenever $T_{n,1}>t_{\alpha;n,q,\btheta_0}=(nh^q)^{-1}\lambda_q(L^2)\lambda_q(L)^{-2}+h^\frac{q}{2}\nu_d\sqrt{2R(f_{\btheta_0})}z_\alpha$. Under local Pitman alternatives of the kind $H_{1P}: f=f_{\btheta_0} + (nh^\frac{q}{2})^\frac{1}{2}\Delta$ ($\Delta=0$ gives $H_0$), where $\Delta:\Om{q}\rightarrow\R$ is such that $\Iq{\Delta(\bx)}{\bx}=0$, and if $\hat\btheta-\bthe_0=\mathcal{O}_\mathbb{P}\big(n^{-\frac{1}{2}}\big)$ under $H_{1P}$, the test rejects if $T_{n,1}>t_{\alpha;n,q,\btheta_0}-R(\Delta)$. Hence, the larger the $L^2$-norm of $\Delta$, the larger the~power. \\

With $f$ being a directional-linear density, testing \eqref{gpcgm:eq:testdensity} can be done using
\begin{align*}
T_{n,2}=&\Iqr{\lrp{\hat f_{h,g}(\bx,y)-LK_{h,g}f_{\hat \btheta}(\bx,y)}^2}{\bx}{y},
\end{align*}
where $LK_{h,g} f_{\btheta_0}(\bx,y)=\Iqr{LK_{h,g}\lrp{(\bx,y),(\bz,t)}f_{\btheta_0}(\bz,t)}{\bz}{t}$ is the expected value of $\hat f_{h,g}(\bx,y)$ under $H_0$. Under regularity assumptions for the density and kernels (A1, A2 and A5 in \cite{Garcia-Portugues:clt}), and $\hat\btheta-\bthe_0=\mathcal{O}_\mathbb{P}\big(n^{-\frac{1}{2}}\big)$ under $H_{1P}: f=f_{\btheta_0} + (nh^\frac{q}{2})^\frac{1}{2}\Delta$ ($\Delta:\Om{q}\times\R\rightarrow\R$ is such that $\Iqr{\Delta(\bx,y)}{\bx}{y}=0$), the limit law of $T_{n,2}$ under $H_{1P}$ is
\begin{align}
n(h^qg)^\frac{1}{2}\lrp{T_{n,2}-\frac{\lambda_q(L^2)\lambda_q(L)^{-2}R(K)}{nh^qg}}\stackrel{d}{\longrightarrow} \mathcal{N}\lrp{R(\Delta),2\nu_d^2\nu_l^2R(f_{\btheta_0})},\label{gpcgm:eq:tn2}
\end{align}
where $\nu_l^2=\Ir{\lrb{\Ir{K(u)K(u+v)}{u}}^2}{v}$. $\nu_d^2$ and $\nu_l^2$ are the variance components associated to the smoothing and, for the Gaussian and von Mises kernels, their expressions are remarkably simple: $\nu_l^2=(8\pi)^{-\frac{1}{2}}$ and $\nu_d^2=(8\pi)^{-\frac{q}{2}}$. \\

Estimator \eqref{gpcgm:eq:kdedl} allows also to check the independence between the rv's $\bX$ and $Y$ in an omnibus way, for arbitrary dimensions. This degree of generality contrasts with the available tests for assessing the independence between directional and linear variables, mostly focused on the circular case and on the examination of association coefficients (e.g. \cite{Mardia1976}, \cite{Liddell1978}, and \cite{Fisher1981}). Independence can be tested \textit{\`{a} la} \cite{Rosenblatt1975} by considering the problem
\begin{align}
H_0:\; f_{\bX,Y}=f_{\bX}f_{Y}\quad\mbox{vs.}\quad H_1:\; f_{\bX,Y}\neq f_{\bX}f_{Y},\label{gpcgm:eq:testindep}
\end{align}
where $f_{\bX,Y}$ is the joint directional-linear density, and $f_{\bX}$ and $f_{Y}$ are the marginals. To that aim, \cite{Garcia-Portugues:testindep} propose the statistic
\[
T_{n,3}=\Iqr{\lrp{\hat f_{h,g}(\bx,y)-\hat f_{h}(\bx)\hat f_{g}(y)}^2}{\bx}{y}.
\]
Under the same conditions on the density and kernels required for \eqref{gpcgm:eq:tn2}, and with the additional bandwidths' bond $h^{q}g^{-1}\to c$, $0<c<\infty$, the asymptotic distribution of $T_{n,2}$ under independence~is
\begin{align}
n(h^qg)^\frac{1}{2}\lrp{T_{n,3}-A_n}\stackrel{d}{\longrightarrow}\mathcal{N}\lrp{0,2\nu_d^2\nu_l^2R(f_\bX)R(f_Y)},\label{gpcgm:eq:tn3}
\end{align}
where $A_n=\frac{\lambda_{q}(L^2)\lambda_{q}(L)^{-2}R(K)}{nh^qg}-\frac{\lambda_{q}(L^2)\lambda_{q}(L)^{-2}R(f_Y)}{nh^q}-\frac{R(K)R(f_\bX)}{ng}$. Note that \eqref{gpcgm:eq:tn3} is similar to \eqref{gpcgm:eq:tn2}, plus two extra bias terms given by the marginal KDEs. \\

$T_{n,2}$ and $T_{n,3}$ can be modified to work with a directional-directional rv by using the KDE in (\ref{gpcgm:eq:kdedd}). The statistics for \eqref{gpcgm:eq:testdensity} and  \eqref{gpcgm:eq:testindep} are now:
\begin{align*}
T_{n,4}=&\,\Iqq{\lrp{\hat f_{h_1,h_2}(\bx_1,\bx_2)-LK_{h_1,h_2}f_{\hat \btheta}(\bx_1,\bx_2)}^2}{\bx_1}{\bx_2},\\
T_{n,5}=&\,\Iqq{\lrp{\hat f_{h_1,h_2}(\bx_1,\bx_2)-\hat f_{h_1}(\bx_1)\hat f_{h_2}(\bx_2)}^2}{\bx_1}{\bx_2},
\end{align*}
respectively. Under the directional-directional analogues of the assumptions required for \eqref{gpcgm:eq:tn2} and \eqref{gpcgm:eq:tn3}, the asymptotic rejection rule of $T_{n,4}$ is
$T_{n,4}>(nh_1^{q_1}h_2^{q_2})^{-1}\lambda_{q_1}(L^2)\lambda_{q_2}(L^2)(\lambda_{q_1}(L)\lambda_{q_2}(L)^{-2}+(h_1^{q_1}h_2^{q_2})^\frac{1}{2}\nu_{d_1}\nu_{d_2}\sqrt{2R(f_{\btheta_0})}z_\alpha$ and, under independence,  $$
n(h_1^{q_1}h_2^{q_2})^\frac{1}{2}\lrp{T_{n,5}-B_n}\stackrel{d}{\longrightarrow}\mathcal{N}\lrp{0,2\nu_{d_1}^2\nu_{d_2}^2R(f_{\bX_1})R(f_{\bX_2})},
$$  
with $B_n=\frac{\lambda_{q_1}(L^2)\lambda_{q_1}(L)^{-2}\lambda_{q_2}(L^2)\lambda_{q_2}(L)^{-2}}{nh_1^{q_1}h_2^{q_2}}-\frac{\lambda_{q_1}(L^2)\lambda_{q_1}(L)^{-2}R(f_{\bX_2})}{nh_1^{q_1}}-\allowbreak\frac{\lambda_{q_2}(L^2)\lambda_{q_2}(L)^{-2}R(f_{\bX_1})}{nh_2^{q_2}}$.

\subsection{Regression-based tests}

The testing of \eqref{gpcgm:eq:testregression} (\textit{i.e.}, the assessment of whether $m$ has a parametric structure $m_{\btheta_0}$, with $\btheta_0$ either specified or unspecified) is rooted on the nonparametric estimator for $m$ introduced in \eqref{gpcgm:eq:mhp}. In a similar way to \cite{Hardle1993} in the linear setting, problem \eqref{gpcgm:eq:testregression} may be approached with the test statistic
\[
T_{n,6}=\Iq{\lrp{\hat m_{h,p}(\bx)-\mathcal{L}_{h,p}m_{\hat\bthe}(\bx) }^2\hat f_h(\bx)w(\bx)}{\bx},
\]
where $\mathcal{L}_{h,p}m_{\bthe_0}(\bx)=\sum_{i=1}^n W_{n,i}^p\lrp{\bx}m_{\bthe_0}(\bX_i)$ is the smoothing of $m_{\bthe_0}$, included to reduce the asymptotic bias \citep{Hardle1993}, and $w:\Om{q}\rightarrow\R_0^+$ is an optional weight function. The inclusion of $\hat f_h$ has the benefits of avoiding the presence of the density of $\bX$ in the asymptotic bias and variance, and of mitigating the effects of the squared difference in sparse areas of $\bX$. \\\noclub[4]

Under $H_0$, $\hat\btheta-\bthe_0=\mathcal{O}_\mathbb{P}\big(n^{-\frac{1}{2}}\big)$, and certain  regularity conditions (A1--A3 and A5 in \cite{Garcia-Portugues:locgof}), the limit distribution of $T_{n,6}$ \nolinebreak[4]is
\[
nh^\frac{q}{2}\lrp{T_{n,6}-\frac{\lambda_q(L^2)\lambda_q(L)^{-2}}{nh^{q}}\Iq{\sigma_{\bthe_0}^2(\bx)w(\bx)}{\bx}}\stackrel{d}{\longrightarrow} \mathcal{N}\lrp{0,2\nu_d^2R\lrp{\sigma_{\bthe_0}^2w}},
\]
where $\sigma_{\bthe_0}^2(\bx)=\Esp{(Y-m_{\bthe_0}(\bX))^2\vert \bX=\bx}$, this is, $\V{Y\vert\bX=\bx}$ under $H_0$.

\section{Convergence towards the asymptotic distribution}
\label{gpcgm:sec:boot}

Unfortunately, the asymptotic distributions of the test statistics $T_{n,k}$, $k=1,\ldots,6$ are almost useless in practise. In addition to the unknown quantities present in the asymptotic distributions, the convergences toward the limits are slow and depend on the bandwidth sequences. This forces the consideration of resampling mechanisms for calibrating the distributions of the statistics under the null: parametric bootstraps in $T_{n,1}$, $T_{n,2}$, and $T_{n,4}$ \citep{Boente2013,Garcia-Portugues:clt}; a \textit{wild} bootstrap for $T_{n,6}$ \citep{Garcia-Portugues:locgof}; and a permutation approach for $T_{n,3}$ and $T_{n,5}$ \citep{Garcia-Portugues:testindep}. The purpose of this section is to illustrate, as an example, the convergence to the asymptotic distribution of the statistics $T_{n,3}$ and $T_{n,6}$ via insightful numerical experiments.

\begin{figure}[htpb]
	\centering
	\vspace*{-0.75cm}
	\includegraphics[width=0.525\textwidth]{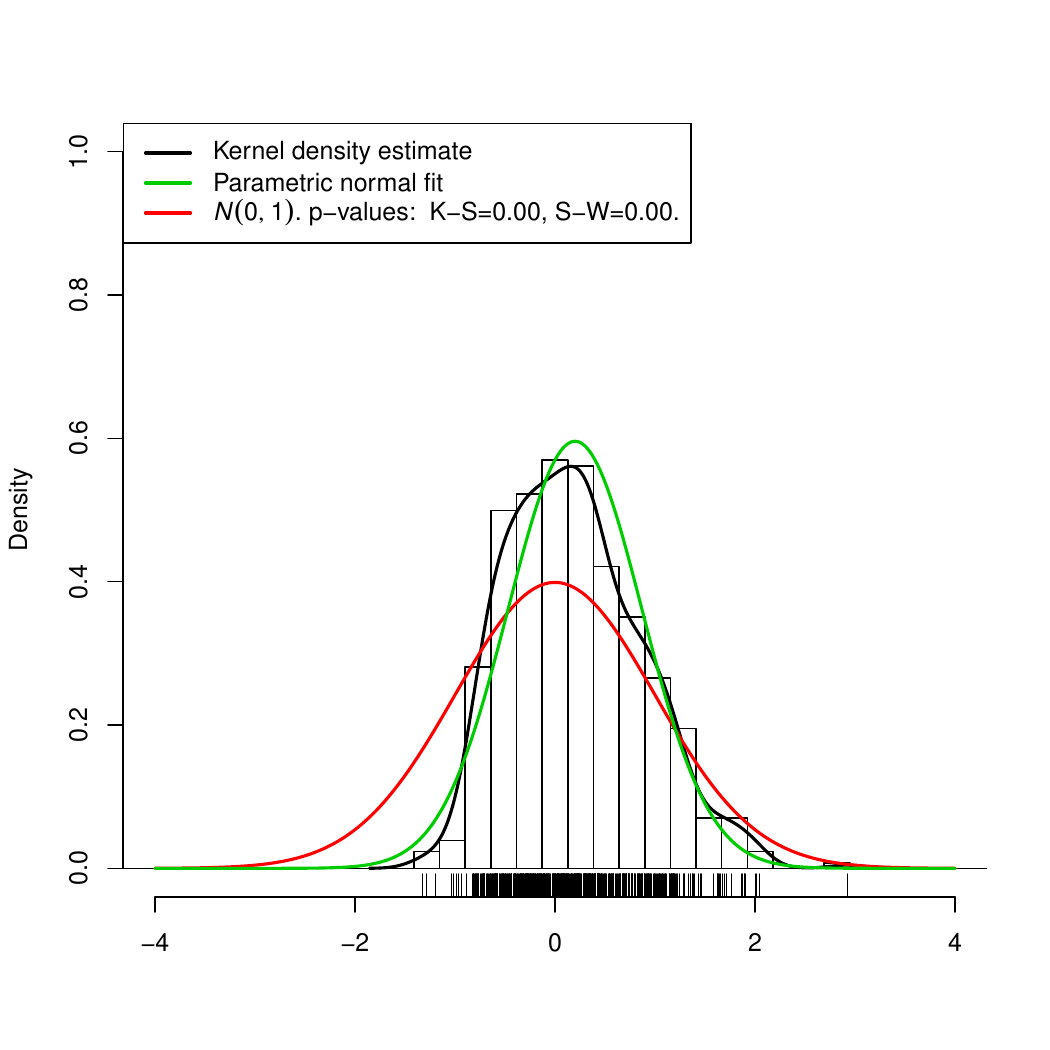}\hspace*{-0.05\textwidth}%
	\includegraphics[width=0.525\textwidth]{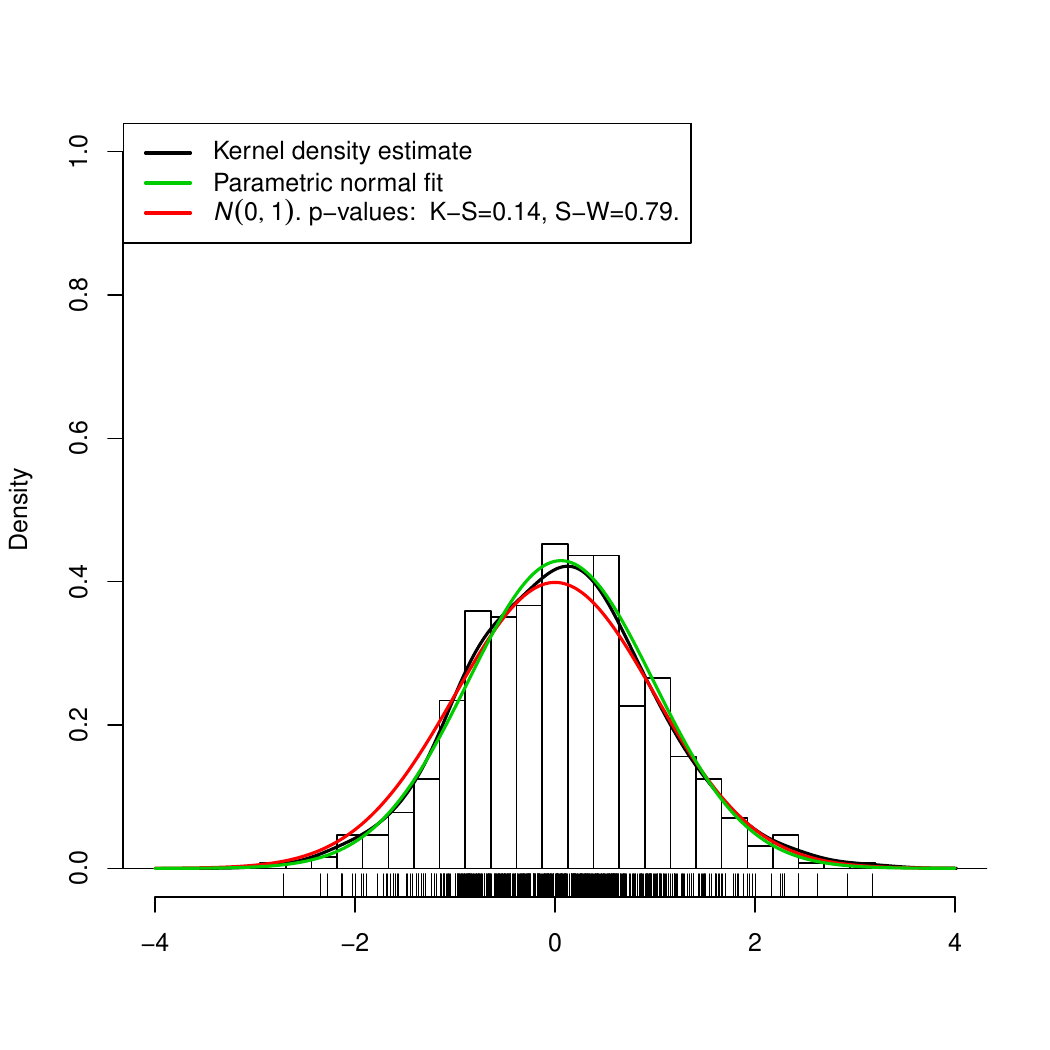}
	\vspace{-0.5cm}
	\caption{\small Asymptotic and empirical distributions for the standardized statistic $T_{n,3}$, for sample sizes $n=10^3$ (left) and $n=5\times10^5$ (right).\label{gpcgm:fig:1}}
\end{figure}

First, for $T_{n,3}$ we considered a circular-linear framework ($q=1$), with a von Mises density with mean $\boldsymbol{\mu}=(0,1)$ and concentration $\kappa=1$ for the circular variable, and a $\mathcal{N}(0,1)$ for the linear density. We also took von Mises and normal kernels. These choices gave $R(f_\bX)=(2\pi)^{-1}\mathcal{I}_{0}(2) \mathcal{I}_{0}(1)^{-2}$ ($\mathcal{I}_{0}$ stands for the modified Bessel function of first kind and order $0$), $R(f_Y)=\big(2\pi^\frac{1}{2}\big)^{-1}$, $\nu_{d}^2=\nu_l^2=(8\pi)^{-\frac{1}{2}}$, and $R(K)=\lambda_1(L^2)\lambda_1(L)^{-2}=\big(2\pi^\frac{1}{2}\big)^{-1}$. We simulated $M=500$ samples of size $n=5^k\times10^l$, $k=0,1$, $l=1,\ldots,6$ under independence, obtaining $\big\{n\big(\frac{h_ng_n}{2\nu_d^2\nu_l^2R(f_\bX)R(f_Y)}\big)^\frac{1}{2}\big(T^j_{n,3}-A_n\big)\big\}_{j=1}^M$. We took $h_n=g_n=2n^{-\frac{1}{3}}$ as a compromise between fast convergence and avoiding numerical instabilities. Figure \ref{gpcgm:fig:1} shows several density estimates for the sample of standardized statistics, jointly with the $p$-values of the Kolmogorov--Smirnov (K--S) test for $\mathcal{N}(0,1)$, and of the Shapiro--Wilk (S--W) test for normality. Both tests are significant up to a very large sample size (close to $n=5\times10^5$ data), which is apparent from the visual disagreement between the finite sample and asymptotic distributions for $n=10^3$. \\\nowidow[3]

\begin{figure}[h]
	\centering
	\vspace{-0.75cm}
	\includegraphics[width=0.525\textwidth]{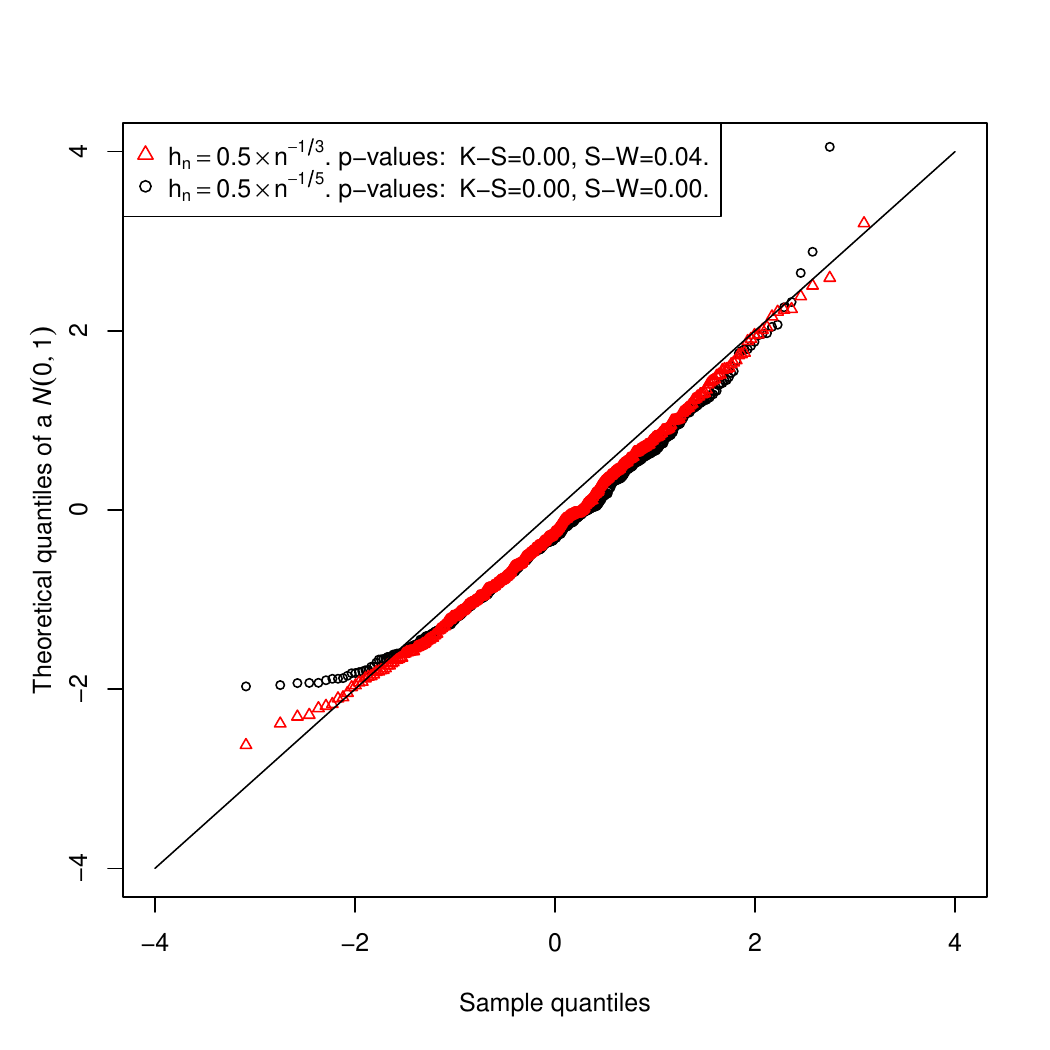}\hspace*{-0.025\textwidth}\includegraphics[width=0.525\textwidth]{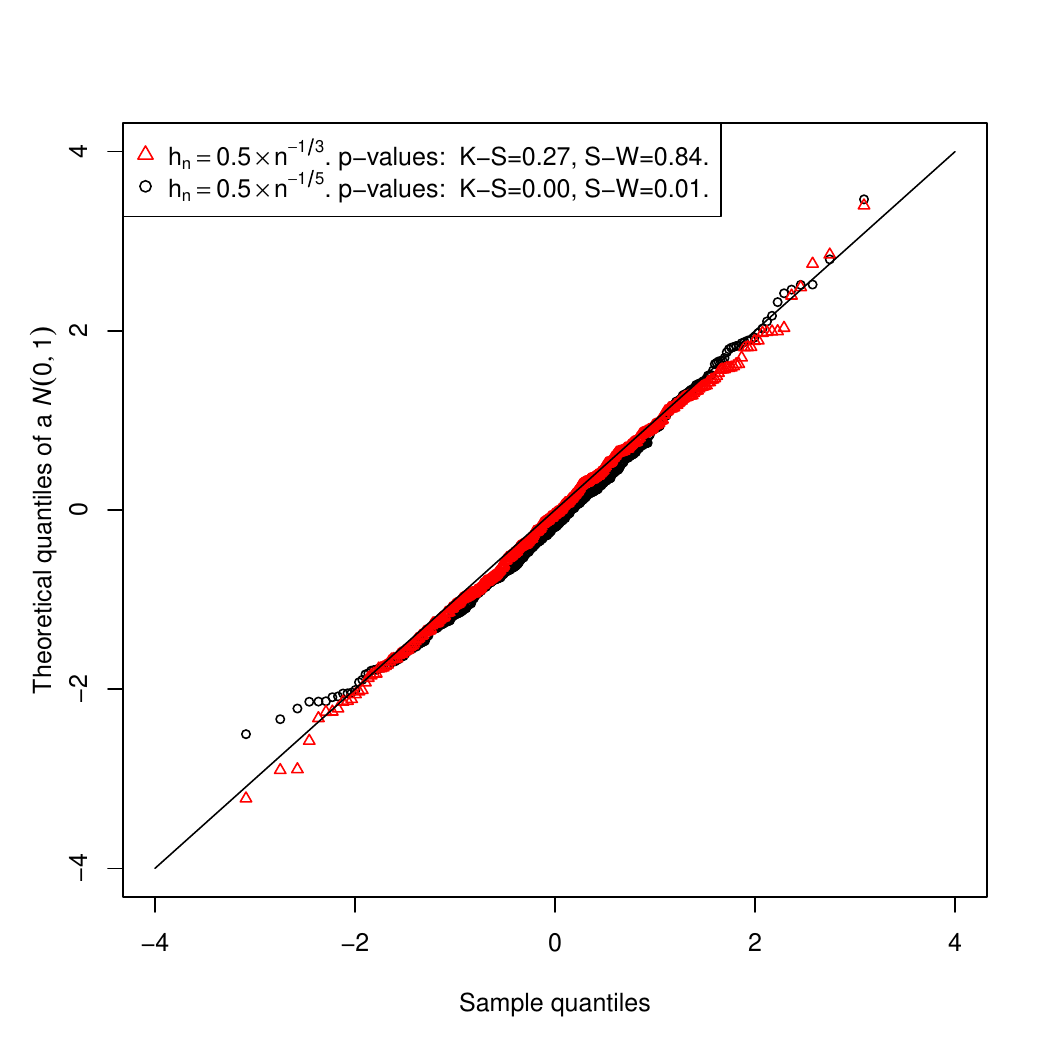}
	\vspace{-0.5cm}
	\caption{\small QQ-plot comparing the sample quantiles of $\big\{nh^\frac{1}{2}\big(\frac{128}{\pi}\big)^{\frac{1}{4}}\big(T^j_n-\frac{\sqrt{\pi}}{4}nh\big)\big\}_{j=1}^{M}$ with the ones of the asymptotic distribution, for $n=10^3$ (left) and $n=5\times 10^5$ (right).\label{gpcgm:fig:2}}
\end{figure}

Second, for $T_{n,6}$, the regression model $Y=1+\varepsilon$ is considered, with $\varepsilon\sim\mathcal{N}\lrp{0,\frac{1}{4}}$, and $\mathbf{X}$ uniformly distributed on the circle. The composite hypothesis is $H_0: m\equiv c$, for $c\in\mathbb{R}$ unknown. $H_0$ is checked using the local constant estimator with von Mises kernel and $w\equiv1$. Figure \ref{gpcgm:fig:2} shows the QQ-plots computed from the sample $\big\{nh^\frac{1}{2}\big(\frac{128}{\pi}\big)^{\frac{1}{4}}\big(T^j_n-\frac{\sqrt{\pi}}{4}nh\big)\big\}_{j=1}^{M}$, for the bandwidth sequences $h_n=\frac{n^{-r}}{2}$, $r=\frac{1}{3},\frac{1}{5}$, which were chosen in order to illustrate their impact in the convergence to the asymptotic distribution. Specifically, it can be seen that the effect of undersmoothing boosts the convergence since the bias is mitigated. Again, up to large sample sizes, the degree of disagreement between the finite sample and the asymptotic distributions is quite evident. 

\section*{Acknowledgements}

The authors acknowledge the support of project MTM2016-76969-P from the Spanish State Research Agency (AEI), Spanish Ministry of Economy, Industry and Competitiveness, and European Regional Development Fund (ERDF). We also thank Eduardo Gil, Juan J. Gil, and Mar\'ia Angeles Gil for inviting us to contribute to this volume, in memory of Pedro.


\begin{thebibliography}{}
	
	\bibitem[Bai et~al., 1988]{Bai1988}
	Bai, Z.~D., Rao, C.~R., and Zhao, L.~C. (1988).
	\newblock Kernel estimators of density function of directional data.
	\newblock {\em J. Multivariate Anal.}, 27(1):24--39.
	
	\bibitem[Bickel and Rosenblatt, 1973]{Bickel1973}
	Bickel, P.~J. and Rosenblatt, M. (1973).
	\newblock On some global measures of the deviations of density function
	estimates.
	\newblock {\em Ann. Statist.}, 1(6):1071--1095.
	
	\bibitem[Boente et~al., 2014]{Boente2013}
	Boente, G., Rodr\'iguez, D., and Gonz\'alez-Manteiga, W. (2014).
	\newblock Goodness-of-fit test for directional data.
	\newblock {\em Scand. J. Stat.}, 41(1):259--275.
	
	\bibitem[Durbin, 1973]{Durbin1973}
	Durbin, J. (1973).
	\newblock Weak convergence of the sample distribution function when parameters
	are estimated.
	\newblock {\em Ann. Statist.}, 1:279--290.
	
	\bibitem[Elderton, 1902]{Elderton1902}
	Elderton, W.~P. (1902).
	\newblock Tables for testing the goodness of fit of theory to observation.
	\newblock {\em Biometrika}, 1(2):155--163.
	
	\bibitem[Fan and Gijbels, 1996]{Fan1996}
	Fan, J. and Gijbels, I. (1996).
	\newblock {\em Local polynomial modelling and its applications}, volume~66 of
	{\em Monographs on Statistics and Applied Probability}.
	\newblock Chapman \& Hall, London.
	
	\bibitem[Fan, 1994]{Fan1994}
	Fan, Y. (1994).
	\newblock Testing the goodness of fit of a parametric density function by
	kernel method.
	\newblock {\em Economet. Theor.}, 10(2):316--356.
	
	\bibitem[Fisher and Lee, 1981]{Fisher1981}
	Fisher, N.~I. and Lee, A.~J. (1981).
	\newblock Nonparametric measures of angular-linear association.
	\newblock {\em Biometrika}, 68(3):629--636.
	
	\bibitem[Garc\'ia-Portugu\'es et~al., 2014]{Garcia-Portugues:testindep}
	Garc\'ia-Portugu\'es, E., Barros, A. M.~G., Crujeiras, R.~M.,
	Gonz\'alez-Manteiga, W., and Pereira, J. (2014).
	\newblock A test for directional-linear independence, with applications to
	wildfire orientation and size.
	\newblock {\em Stoch. Environ. Res. Risk Assess.}, 28(5):1261--1275.
	
	\bibitem[Garc\'ia-Portugu\'es et~al., 2013]{Garcia-Portugues:dirlin}
	Garc\'ia-Portugu\'es, E., Crujeiras, R.~M., and Gonz\'alez-Manteiga, W. (2013).
	\newblock Kernel density estimation for directional-linear data.
	\newblock {\em J. Multivariate Anal.}, 121:152--175.
	
	\bibitem[Garc\'ia-Portugu\'es et~al., 2015]{Garcia-Portugues:clt}
	Garc\'ia-Portugu\'es, E., Crujeiras, R.~M., and Gonz\'alez-Manteiga, W. (2015).
	\newblock Central limit theorems for directional and linear data with
	applications.
	\newblock {\em Statist. Sinica}, 25:1207--1229.
	
	\bibitem[Garc\'ia-Portugu\'es et~al., 2016]{Garcia-Portugues:locgof}
	Garc\'ia-Portugu\'es, E., Van~Keilegom, I., Crujeiras, R., and
	Gonz\'alez-Manteiga, W. (2016).
	\newblock Testing parametric models in linear-directional regression.
	\newblock {\em Scand. J. Statist.}, 43(4):1178--1191.
	
	\bibitem[Gonz\'alez-Manteiga and Crujeiras, 2013]{Gonzalez-Manteiga2013}
	Gonz\'alez-Manteiga, W. and Crujeiras, R.~M. (2013).
	\newblock An updated review of goodness-of-fit tests for regression models.
	\newblock {\em Test}, 22(3):361--411.
	
	\bibitem[Hall et~al., 1987]{Hall1987}
	Hall, P., Watson, G.~S., and Cabrera, J. (1987).
	\newblock Kernel density estimation with spherical data.
	\newblock {\em Biometrika}, 74(4):751--762.
	
	\bibitem[H{\"a}rdle and Mammen, 1993]{Hardle1993}
	H{\"a}rdle, W. and Mammen, E. (1993).
	\newblock Comparing nonparametric versus parametric regression fits.
	\newblock {\em Ann. Statist.}, 21(4):1926--1947.
	
	\bibitem[Liddell and Ord, 1978]{Liddell1978}
	Liddell, I.~G. and Ord, J.~K. (1978).
	\newblock Linear-circular correlation coefficients: some further results.
	\newblock {\em Biometrika}, 65(2):448--450.
	
	\bibitem[Mardia, 1976]{Mardia1976}
	Mardia, K.~V. (1976).
	\newblock {Linear-circular correlation coefficients and rhythmometry}.
	\newblock {\em Biometrika}, 63(2):403--405.
	
	\bibitem[Mardia and Jupp, 2000]{Mardia2000}
	Mardia, K.~V. and Jupp, P.~E. (2000).
	\newblock {\em Directional statistics}.
	\newblock Wiley Series in Probability and Statistics. John Wiley \& Sons,
	Chichester, second edition.
	
	\bibitem[Pearson, 1900]{Pearson1900}
	Pearson, K. (1900).
	\newblock On the criterion that a given system of deviations from the probable
	in the case of a correlated system of variables is such that it can be
	reasonably supposed to have arisen from random sampling.
	\newblock {\em Philos. Mag. Series 5}, 50(302):157--175.
	
	\bibitem[Pearson, 1916]{Pearson1916}
	Pearson, K. (1916).
	\newblock On the application of ``goodness of fit'' tables to test regression
	curves and theoretical curves used to describe observational or experimental
	data.
	\newblock {\em Biometrika}, 11(3):239--261.
	
	\bibitem[Rosenblatt, 1975]{Rosenblatt1975}
	Rosenblatt, M. (1975).
	\newblock A quadratic measure of deviation of two-dimensional density estimates
	and a test of independence.
	\newblock {\em Ann. Statist.}, 3(1):1--14.
	
	\bibitem[Zhao and Wu, 2001]{Zhao2001}
	Zhao, L. and Wu, C. (2001).
	\newblock Central limit theorem for integrated square error of kernel
	estimators of spherical density.
	\newblock {\em Sci. China Ser. A}, 44(4):474--483.
	
\end{thebibliography}

\end{document}